\documentclass[sigconf]{acmart}
\usepackage{subcaption}
\usepackage{amsthm}
\usepackage{bm}
\usepackage{amsfonts}
\usepackage{multirow}
\usepackage{booktabs}
\usepackage{epstopdf}
\usepackage{amsmath}
\usepackage{algorithm}
\usepackage{xspace}
\usepackage{array}
\usepackage{enumitem} 
\usepackage{multirow}

\AtBeginDocument{%
  \providecommand\BibTeX{{%
    \normalfont B\kern-0.5em{\scshape i\kern-0.25em b}\kern-0.8em\TeX}}}

\newcommand{\eat}[1]{}

\newcommand{\ie}{\emph{i.e.,}\xspace}
\newcommand{\eg}{\emph{e.g.,}\xspace}
\newcommand{\etal}{\emph{et al.}\xspace}
\newcommand{\baby}{\textsc{PDN}\xspace}

\copyrightyear{2021}
\acmYear{2021}
\setcopyright{acmcopyright}\acmConference[SIGIR '21]{Proceedings of the 44th International ACM SIGIR Conference on Research and Development in Information Retrieval}{July 11--15, 2021}{Virtual Event, Canada}
\acmBooktitle{Proceedings of the 44th International ACM SIGIR Conference on Research and Development in Information Retrieval (SIGIR '21), July 11--15, 2021, Virtual Event, Canada}
\acmPrice{15.00}
\acmDOI{10.1145/3404835.3462878}
\acmISBN{978-1-4503-8037-9/21/07}

\begin{document}
\fancyhead{}
\title{Path-based Deep Network for Candidate Item Matching in Recommenders}\thanks{$\dagger$ Chenliang Li and Hongbo Deng are the corresponding authors.}
\author{Houyi Li$^{1}$, Zhihong Chen$^{1}$, Chenliang Li$^{3 \dagger}$, Rong Xiao$^{1}$, Hongbo Deng$^{1 \dagger}$, Peng Zhang$^1$, \\ Yongchao Liu$^{2}$, Haihong Tang$^1$}
\affiliation{
	\institution{
		$^1$Alibaba Group, Hangzhou, China, \qquad $^2$Ant Group, Hangzhou, China\\
		$^3$School of Cyber Science and Engineering, Wuhan University, Wuhan, China\\
		$^1$\{houyi.lhy, jhon.czh, xiaorong.xr, dhb167148, zhangpeng04, piaoxue\}@alibaba-inc.com\\
		$^3$cllee@whu.edu.cn, $^2$yongchao.ly@antgroup.com
	}
	\country{}
}
\def\authors{Houyi Li, Zhihong Chen, Chenliang Li, Rong Xiao, Hongbo Deng, Peng Zhang, Yongchao Liu, Haihong Tang}



\begin{abstract}
The large-scale recommender system mainly consists of two stages: matching and ranking. 
The matching stage (also known as the retrieval step) identifies a small fraction of relevant items from billion-scale item corpus in low latency and computational cost. Item-to-item collaborative filtering (item-based CF) and embedding-based retrieval (EBR) have been long used in the industrial matching stage owing to its efficiency. 
However, item-based CF is hard to meet personalization, while EBR has difficulty in satisfying diversity. In this paper, we propose a novel matching architecture, Path-based Deep Network (named \baby), through incorporating both personalization and diversity to enhance matching performance. Specifically, \baby is comprised of two modules: \textit{Trigger Net} and \textit{Similarity Net}. \baby utilizes Trigger Net to capture the user's interest in each of his/her interacted item. Similarity Net is devised to evaluate the similarity between each interacted item and the target item based on these items' profile and CF information. The final relevance between the user and the target item is calculated by explicitly considering user's diverse interests, \ie aggregating the relevance weights of the related two-hop paths (one hop of a path corresponds to user-item interaction and the other to item-item relevance). Furthermore, we describe the architecture design of the proposed \baby in a leading real-world E-Commerce service (Mobile Taobao App). Based on offline evaluations and online A/B test, we show that \baby outperforms the existing solutions for the same task. The online results also demonstrate that \baby can retrieve more personalized and more diverse items to significantly improve user engagement. Currently, \baby system has been successfully deployed at Mobile Taobao App and handling major online traffic.
\end{abstract}




\maketitle

\section{Introduction}
Recommender systems are important in customer-oriented E- Commerce platforms (e.g., Taobao and Amazon). The purpose is to connect users to their preferred items and produce more profits. Due to the tremendous number of items available in the E-Commerce platforms, many industrial systems~\cite{YoutBednn,gomez2015netflix} are devised with a matching stage and a ranking stage. Specifically, a matching stage is expected to retrieve a small fraction of relevant items in low latency and computational cost, and a ranking stage aims to refine the ranking of these relevant items in terms of the user's interest with more complex models. In this work, we focus on the matching stage since it is the fundamental part and also the bottleneck of the system.

Item-to-item based collaborative filtering, as known as item-based CF, is an information retrieval solution for item matching. It estimates the relevance between two items based on their co-occurrence patterns. Several outstanding merits make item-based CF a natural choice for the matching stage in industry~\cite{davidson2010youtube,liu2017related,smith2017two}: (1) Since only the items previously interacted by the user are used for retrieval, an item-based CF could well match the efficiency requirement for many online services; (2) Furthermore, a user's interest could be very diverse. By taking each interacted item into consideration, the user's interest can be well covered to its maximum; (3) At last, the relevance between items is mainly derived based on massive user behaviors. These signals are effective to identify the relevant items w.r.t the interacted ones.

However, this kinds of methods also have some limitations. The traditional inverted indexes are hard to meet subtle personalization needs~\cite{zhang2020towards}. By considering only item co-occurrence patterns, item-based CF is inferior to accommodate the user's unique characteristics, \eg gender, consumption capacity and others. Similarly, given the volume of items available in a E-Commerce platform keeps growing, without considering auxiliary information, item-based CF could suffer a lot from the data sparsity problem~\cite{li2019multi}.

To overcome the above limitations, embedding-based retrieval (EBR), especially the deep learning networks with a two-tower architecture, has drawn growing interests recently~\cite{huang2020embedding,yi2019sampling,yang2020mixed}. Briefly speaking, EBR aims to represent each user and item by embedding their profile respectively. In this sense, the matching process is transferred to perform nearest neighbor (NN) search in the embedding space. Although the representation learning could alleviate the data sparsity problem to some extent, these methods also have some limitations. The two-tower architecture is not easy to explicitly integrate the co-occurrence information between items. Moreover, a user is often represented as a single embedding vector, which is insufficient to encode the diversity of the user's interest~\cite{DSSM,YoutBednn}.

A normal user would interact with hundreds of items that belong to different categories every month, indicating the diversity of user interests. Actually, in real industrial systems, to simultaneously capture the diversity of user interests and ensure personalization, typically there are multiple strategies, e.g., various kinds of inverted indexes based on collaborative filtering models and EBR strategies with different network structures. These models are deployed in parallel for the matching task. Note that the candidates are usually generated with the relevance scores in different scales. It is not straightforward to fuse these incomparable values for promising performance. We argue that this multi-strategy solution may be suboptimal due to the high maintenance cost of the these strategies and lack of tailored joint optimization. 

\begin{figure}[t]
  	\centering
  	\includegraphics[width=0.82\linewidth]{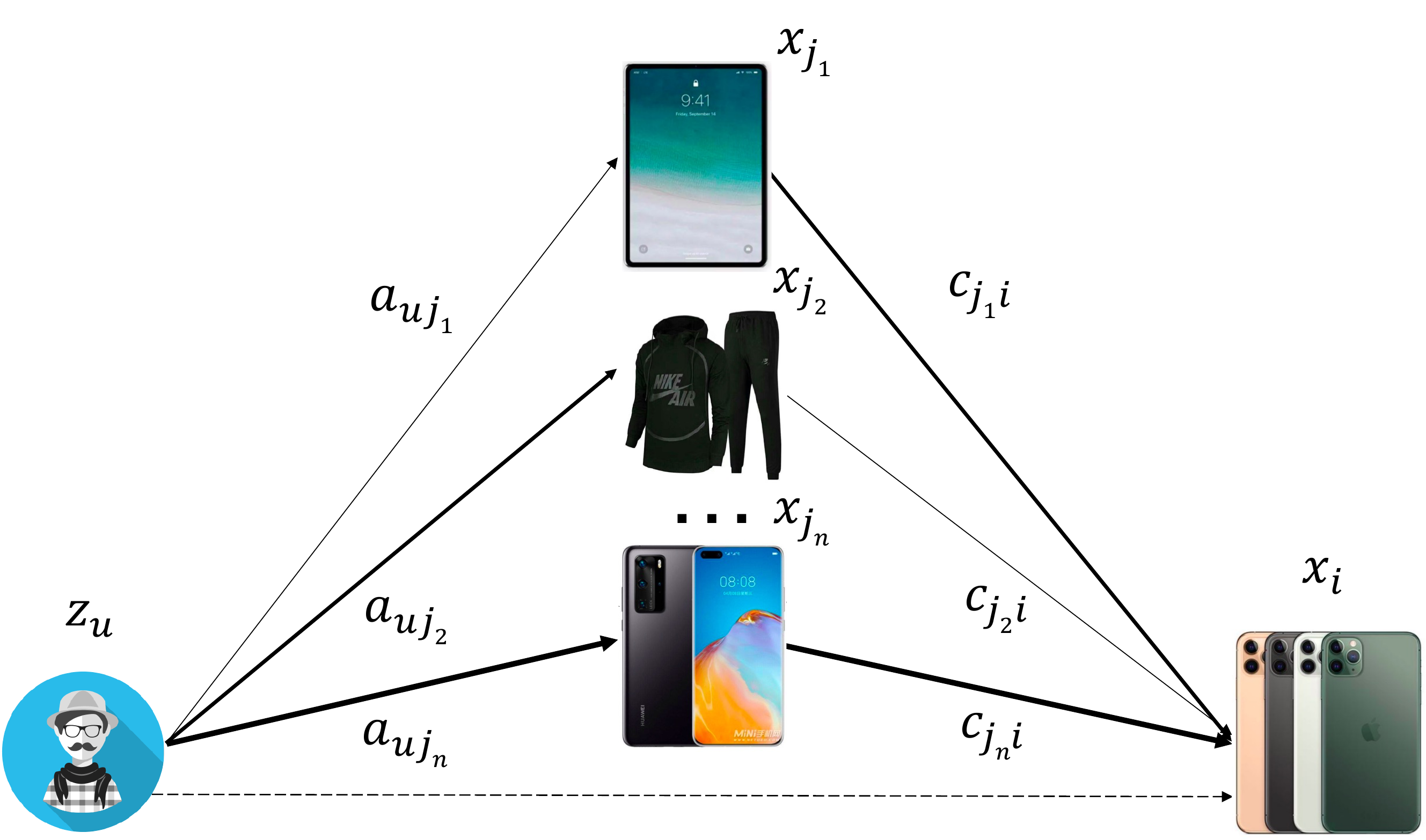}
 	\caption{Modeling the preference of user $u$ to target item $i$ with a 2-hop graph where $j_1$ to $j_n$ represent interacted items. The thickness of edge indicates the relevance.}
 	\label{fig:path}
 	\vspace{-0.5cm}
\end{figure}

Deep Interest Network (DIN)~\cite{din} introduces the similarity between interacted and target items through the target attention for a better recommendation. However, this attention mechanism is simply used to fuse user interaction sequences, which ignores the user's interest in each interacted item. Also, DIN is difficult to be applied for the matching stage since it requires recalculating the user representation for each target item. Inspired by DIN, in this work, we propose a novel matching architecture called Path-based Deep Network (\baby), which decouples the target item based attention from user representation learning by building a marriage between item-based CF and EBR. In \baby, we use the thought of representation learning of EBR (\ie user profile, interacted item sequence and item profile) and item-based CF (\ie item co-occurrence) to accommodate both user personalization and diverse interest modeling for better performance. Specifically, \baby consists of two main subnetworks: Trigger Net and Similarity Net. Trigger Net (\textit{TrigNet}) is introduced to encode the user's interest by considering each interacted item as a trigger\footnote{The terms \textit{interacted item, trigger, trigger item} are exchangeable since they refer to the same meaning in this paper.}. That is, the generated user representation has a variable dimension such that each dimension describes the user's interest on the interacted item. Analogous to item-based CF, Similarity Net (\textit{SimNet}) generates an item representation and each dimension describes the similarity between an interacted item and the target item. Note that, the dimensions of user representation and item representation extracted by EBR are constant, while the dimensions of the user and item representations extracted by \baby are variable which are equal to the number of the user's triggers. As shown in Figure~\ref{fig:path}, by connecting the user with the target item through her interacted items, we can form a series of 2-hop paths. Based on these 2-hop paths, \baby aggregate the relevance between the user and the target item by explicitly considering a user's diverse interests for better performance. Another merit is that the whole model is trained with an end-to-end fashion. Therefore, the relevance scores can be compared with each other in a uniform way.

It is worthwhile to highlight that the proposed \baby bears the advantages of both item-based CF and EBR for efficient online processing. On one hand, empowered by the feature embedding, we can utilize \textit{TrigNet} to extract top-$m$ most important triggers w.r.t. the user interests to satisfy the real-time requirement. On the other hand, \textit{SimNet} works independently from \textit{TrigNet}. We can apply parallel computing to support offline index construction with item-to-item relevance calculated by \textit{SimNet} efficiently. In summary, the main contributions of this paper are as follows:

\begin{itemize}
\item \textbf{A novel matching model.} We propose Path-based deep network by incorporating advantages of item-based CF and EBR. \baby integrates all of profile information for user attention and co-occurrence patterns between items for target attention in the form of 2-hop path aggregation. Both the user personalization and the interest diversity are accommodated for better item matching.

\item \textbf{Efficient online retrieval.} We construct an industrial-scale online matching system based on \baby. In particular, we describe how to leverage \baby for item retrieval in low latency and computational cost.

\item \textbf{Offline and online experiments.} Extensive offline experiments on several real-world datasets demonstrates that the proposed \baby achieves much better performance than the existing alternatives. Besides, we evaluate \baby on the recommender system of Taobao with A/B test over a two-week period. The results suggest that a large performance gain is obtained on almost all metrics.
\end{itemize}

\section{Related Work}
CF-based methods are successful in building recommender systems at the matching stage~\cite{su2009survey}. Among them, item-based CF~\cite{sarwar2001item,linden2003amazon}, which calculates the similarity matrix of items in advance, and recommend items similar to the user's clicked ones, has been widely employed in industrial settings due to its interpretability and efficiency. Early works utilize statistical measures such as cosine similarity and Pearson coefficient to estimate item similarities. In recent years, several approaches attempt to learn item similarities by optimizing a recommendation-aware objective function. Ning~\etal~\cite{ning2011slim} propose SLIM for item relevance learning by minimizing the loss between the original user-item interaction matrix and the reconstructed one. He~\etal~\cite{NAIS} propose NAIS with an attention mechanism to distinguish the different importance of historical items in a user profile, which shares a similar idea with DIN~\cite{din}. However, attention mechanism based methods are only applicable to the ranking stage due to the complexity of computation.

With the success of EBR~\cite{bengio2013representation}, 
two-tower architectures based on deep neural networks have been widely adopted in industrial recommender systems to capture personalized information of a user by leveraging rich content features~\cite{huang2020embedding,yi2019sampling,yang2020mixed}. 
Note that, in the matching stage, to process billions or trillions of items in low latency and computational cost, 
the two towers can not interact with each other to ensure the parallel feature extraction. 
Particularly, the DSSM-based model~\cite{DSSM} learns the relevance based on the inner product between user features and item features. To extract more discriminative user features, 
Youtube DNN~\cite{YoutBednn} extends user features by average pooling user behavior, 
while BST~\cite{chen2019behavior} utilizes the powerful Transformer model to capture the sequential signals underlying users' behavior sequences. 

Different from the above methods, the PDN we propose combines the advantages of item-based CF and EBF based on deep neural networks, which constructs user-item subnetwork to ensure personalization similar to EMB, and constructs item-item subnetwork to capture multiple user interests similar to item-based CF.

\section{Preliminaries}

Figure~\ref{fig:path} summarizes the recommendation problem in the form of 2-hop paths between the user and the target item. Here, $\bm{z}_u$ represents the user information, \eg user id, age, gender, click times, purchase times of each category, 
for user $u$; $\bm{x}_i$ represents the information of target item, \eg item id, brand id, category id, monthly sales of target item $i$; $\{\bm{x}_{j_k}\}_{k=1}^n$ represent the auxiliary information of items $\{j_{k}\}_{k=1}^n$ interacted by $u$; $\{\bm{a}_{u j_k}\}_{k=1}^n$ represent the behavior information from user $u$ on these interacted items, \eg stay time, purchase times; and $\{\bm{c}_{j_k i}\}_{k=1}^n$ represent the relevance information between interacted items and target item, which is obtained from an item-based CF algorithm or statistical correlation measure based on item co-occurrence patterns. 

As shown in Figure~\ref{fig:path}, there is a direct link between the user and the target item (shown in dashed line), which indicates the user's intuitive interest on the target item. Also, we can further form $n$ 2-hop paths by bridging through the interacted items. The first hop represents the user's interest in the interacted items and the second hop represents the relevance between the interacted items and the target item. Hence, the item matching with the above available information for recommendation can be formulated as:
\begin{equation}
\hat{y}_{ui} = f\big( \bm{z}_u, \bm{x}_i, \{\bm{x}_{j}\}, \{\bm{a}_{uj}\}, \{\bm{c}_{ji}\} \big) ~ \textit{with}~~ j\in N(u)
\label{eq:general}
\end{equation}
where $f$ is defined as the recommendation algorithm, $\hat{y}_{ui}$ is defined as the relevance score between user $u$ and target item $i$, 
and $N(u)$ is defined as the items interacted by $u$.

Most of the existing work for recommender systems, 
including item-based CF and EBR, can be regarded as a special case of Eq.~\ref{eq:general}. 
For example, the regression form of item-based CF~\cite{itemcf} can be formulated as: 
\begin{equation}
\hat{y}_{ui} = f\big(\{\bm{a}_{uj}\}, \{\bm{c}_{ji}\} \big) = \sum_{j\in N(u)}f_r(\bm{a}_{uj})c_{ji}
\label{eq:itemcf}
\end{equation}
where $f_r:\mathcal{R}^m\rightarrow\mathcal{R}^1$ is a weighting function to capture user's interest for each trigger, $c_{ji} \in \mathcal{R}^1$ represents the relevance between interacted item $j$ and target item $i$ based on item co-occurrence information. Hence, the method can be seen as the sum of weights of all 2-hop paths based on $\{\bm{a}_{uj}\}$ and $\{\bm{c}_{ji}\}$, 
and each path weight can be calculated as $f_r(\bm{a}_{uj})c_{ji}$.

Besides, the methods based on EBR are also a special case of Eq.~\ref{eq:general}. 
For example, the matrix factorization (MF)~\cite{MF2009} can be formulated as: 
\begin{equation}
\hat{y}_{ui} = f\big( \bm{z}_u, \bm{x}_i, \{\bm{x}_{j}\} \big)=\boldsymbol{q}_i\big(\boldsymbol{p}_u+\frac{1}{\sqrt{|N(u)|}}\sum_{j\in{N(u)}}\boldsymbol{q}_j\big)^T
\label{eq:mf}
\end{equation}
where the $\boldsymbol{q}_i$, $\boldsymbol{p}_u$, $\boldsymbol{q}_j$ represent the embedding vector for target item information $\bm{x}_i$, user information $\bm{z}_u$ and interacted item $\{\bm{x}_{j}\}$, respectively. 
MF can be regarded as the sum of the weights of $n+1$ paths. Specifically, the weight of direct path is $\boldsymbol{q}_i\boldsymbol{p}_u$,
and the weight of each 2-hop path is $1/\sqrt{|N(u)|}\cdot\boldsymbol{q}_i\boldsymbol{q}_j$. While YoutubeDNN~\cite{YoutBednn} utilizes deep neural networks as a generalization of matrix factorization, which can be formulated as: 
\begin{equation}
\hat{y}_{ui} = f\big( \bm{z}_u, \bm{x}_i, \{\bm{x}_{j}\})=\boldsymbol{q}_i\Big(\textit{MLP}\big(\boldsymbol{p}_u, \frac{1}{|N(u)|}\sum_{j\in N(u)} \boldsymbol{q}_j\big)\Big)^T
\label{eq:youtubednn}
\end{equation}
Here $MLP$ refers to the multilayer perception. DIN can also be formulated as~\cite{din}:
\begin{equation}
\begin{split}
\hat{y}_{ui} & = f\big( \bm{z}_u, \bm{x}_i, \{\bm{x}_{j}\}, \{\bm{a}_{uj}\} \big) \\
&=\textit{MLP}\Big(\boldsymbol{p}_u, \boldsymbol{q}_i, \sum_{j\in N(u)}\big( \textit{MLP}(\boldsymbol{q}_j, \bm{a}_{uj}, \boldsymbol{q}_i) \odot \boldsymbol{q}_j\big)\Big)
\label{eq:din}
\end{split}
\end{equation}

where $\odot$ represents the element-wise product. Note that DIN~\cite{din} can be considered as a representation of a 2-hop path ($u\rightarrow j\rightarrow i$) in terms of relevance between the target item and each trigger. However, it requires re-calculation of path representation for each target item, making it only applicable for the ranking stage.

To ensure the efficiency of retrieval, item-based CF builds inverted index while EBR applies k-nearest neighbors (KNN) search for online serving. 
However, because both model architectures are constrained by efficiency, 
they cannot make use of all available information in Figure~\ref{fig:path}, 
resulting in suboptimal performance. For example, item-based CF lacks user and item profiles, while EBR lacks the explicit co-occurrence information between items. 
Hence, in this paper, we propose a novel architecture, named \baby, to support both personalized and diversity retrieval with low latency.

\section{Method}

\begin{figure*}[t]
  	\centering
  	\includegraphics[width=0.9\linewidth]{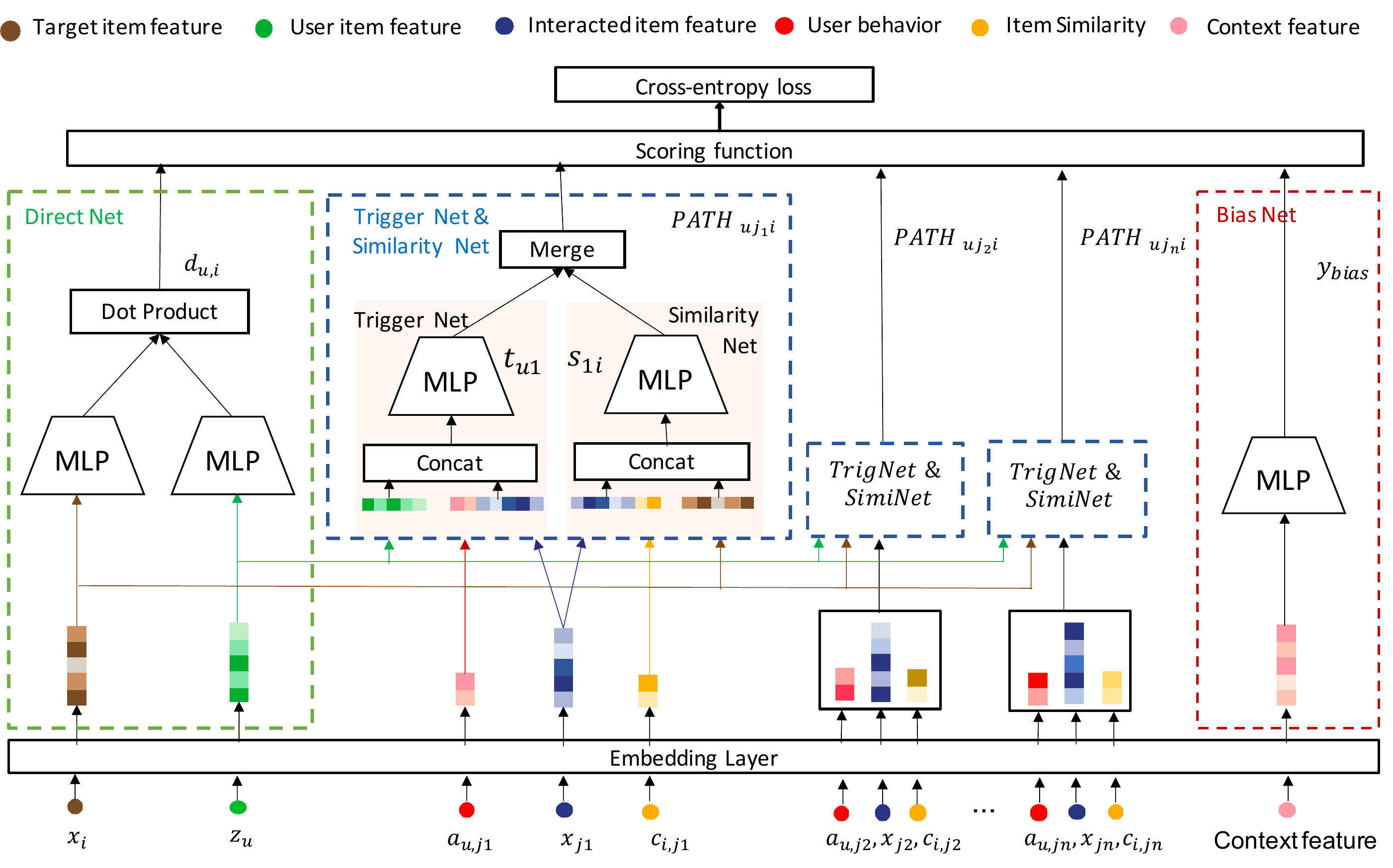}
  	\vspace{-0.2cm}
 	\caption{The network structure of PDN. Direct net is used to obtain the weight of the direct path to capture the user's intuitive interest in the target item, and \textit{TrigNet} and \textit{SimNet} obtain the first-hop weight and the second-hop weight of each 2-hop path respectively to capture the fine-grained user's personalized diversity interests, bias net is used to capture various types of selection bias for further unbiased serving. (Best viewed in color)}
 	\label{fig:pdn_net}
 	\vspace{-0.2cm}
\end{figure*}

In this section, we present the design of the Path-based Deep Network (\baby) for the matching stage of recommender systems. We first introduce the overall architecture of \baby, and then we elaborate on each module of PDN including Embedding Layer, Trigger Net (\textit{TrigNet}), Similarity Net (\textit{SimNet}), Direct Net, and Bias Net. 

\subsection{Overview of PDN}
According to Eq~\ref{eq:general}, the basic workflow of \baby can be formulated as follows:
\begin{subequations}
\begin{equation}
\hat{y}_{ui} = \textit{AGG}\Big(f_{d}\big(\bm{z}_u, \bm{x}_i\big), \big\{\textit{PATH}_{uji}\big\}  \Big) ~ \textit{with}~~ j\in N(u)
\label{eq:pdn_1}
\end{equation}
\begin{equation}
\textit{PATH}_{uji} = \textit{MEG}\big(\textit{TrigNet}(\bm{z}_u, \bm{a}_{uj}, \bm{x}_j), \textit{SimNet}(\bm{x}_j, \bm{c}_{ji}, \bm{x}_i) \big)
\label{eq:pdn_2}
\end{equation}
\label{eq:pdn}
\end{subequations}
where $f_{d}$ is a function to get the relevance weight of the direct path, $\textit{PATH}_{uji}$ represents the relevance weight of the 2-hop path via trigger item $j$,
$\textit{AGG}$ is a scoring function to obtain the final relevance score between the user and the target item by summing relevance weights of $n+1$ paths, 
$\textit{MEG}$ is a function to merge the relevance weights in each 2-hop path,
and $\textit{TrigNet}$, $\textit{SimNet}$ are two independent subnetworks as mentioned previously.

To enable \baby to perform fast online item retrieval, we first define $\textit{MEG}$ as the inner product or summation of the weights of the first hop and the second hop. Then $f_{d}$ is defined also as the inner product between the user and the target item representations. Therefore, the final form of \baby is written as follows: 
\begin{equation}
\hat{y}_{ui} = \boldsymbol{p}_u\boldsymbol{q}_i^T +  \sum_{j\in N(u)}\textit{MEG}\big(\textit{TrigNet}(\bm{z}_u, \bm{a}_{uj}, \bm{x}_j), \textit{SimNet}(\bm{x}_j, \bm{c}_{ji}, \bm{x}_i) \big)
\label{eq:pdn_sum}
\end{equation}

The overall architecture of \baby is shown in Figure~\ref{fig:pdn_net}. 
Note that, in each 2-hop path, when the outputs of \textit{TrigNet} and \textit{SimNet} are vectors, they can be considered as representations of corresponding edges, which is called as vector-based \baby. This setting may have higher model capacity, and make online retrieval more complicated. On the other hand, when the outputs of \textit{TrigNet} and \textit{SimNet} are scalars, they can be regarded as the weights of corresponding edge, which is called as scalar-based \baby.
Comparing with vector-based \baby, scalar-based PDN has lower degree of freedom, which can alleviate the complexity of online retrieval using greedy strategy based on path retrieval. Hence, we introduce each component of \baby with the setting of scalar-based \baby in the following.

\subsection{Feature Composition \& Embedding Layer}
As shown in Figure~\ref{fig:path},  there are four feature fields in our recommender system: user field $\bm{z}_u$, user's behavior field $\{\bm{a}_{uj}\}$, 
item co-occurrence field $\{\bm{c}_{ji}\}$, item field $\bm{x}$ for both of interacted items $\{\bm{x}_{j}\}$ and target item $\bm{x}_{i}$. These fields
include one-hot features, \eg user id, item id, age id, brand id,
and continuous features, \eg monthly sales, stay time, the statistical correlation between items. At first, we transfer dense features into one-hot scheme through the discretization. Then, each one-hot feature is projected into a fixed-size dense representation. After embedding, we concatenate embedding vectors belonging to the same field as the representation of this field.
Formally, field representations of user field, user's behavior field, item co-occurrence field and item field can be written as 
$\bm{E}(\bm{z}_u) \in \mathcal{R}^{1 \times d_u}$, 
$\bm{E}(\bm{a}_{uj}) \in \mathcal{R}^{1 \times d_a}$,
$\bm{E}(\bm{c}_{ji}) \in \mathcal{R}^{1 \times d_c}$,
$\bm{E}(\bm{x}) \in \mathcal{R}^{1 \times d_i}$,  respectively, 
where $d_u$, $d_a$, $d_c$ and $d_i$ are the dimension size of corresponding field respectively. 

\subsection{Trigger Net \& Similarity Net}
After going through the embedding layer, we calculate the relevance weights for each 2-hop path between the user and the target item. For the first hop, we utilize \textit{TrigNet} to capture the user's multi interests by calculating the preference for each of her triggers. Specifically, given user $u$ and her trigger item $j$, the preference score is calculated as follows:
\begin{equation}
t_{uj} = \textit{TrigNet}(\bm{z}_u, \bm{a}_{uj}, \bm{x}_j)=  \textit{MLP}\Big( CAT\big(\bm{E}(z_u), \bm{E}(\bm{a}_{uj}), \bm{E}(\bm{x}_{j})\big) \Big)
\label{eq:trig}
\end{equation}
where $CAT\big(\bm{E}(z_u), \bm{E}(\bm{a}_{uj}), \bm{E}(\bm{x}_{j})\big) \in \mathcal{R}^{1 \times (d_u+d_a+d_i)}$ is the concatenation of the user embedding, user behavior embedding and interacted item embedding, 
and $t_{uj}$ represents the user's preference for an interacted item $j$. 
When there are $n$ distinct interacted items, $\bm{T}_u = [t_{u1},t_{u2},...,t_{un}]$ can be considered as a variable dimension representation for user $u$.
EBR-based methods represent user interests by one fixed dimension representation vector, which can be a bottleneck for capturing diverse interests of users~\cite{li2019multi}, because all information about diverse interests of one user is mixed together, causing inaccurate item retrieval for the matching stage. 
Compared with EBR-based solutions that encodes the user representation with a fixed dimension vector, $\bm{T}_u$ explicitly describes the user's preference for each interacted item, which can better represent the user's diverse interests and is more interpretable. 

It is worth mentioning that the \textit{TrigNet} can employ other more powerful neural networks, such as Recurrent Neural Network (RNN) and transformer-based models for user behavior~\cite{vaswani2017attention, chen2019behavior}. 
However, we would like to emphasize that a simple MLP is more cost-effective to our industrial system. 

As to the second hop, we utilize \textit{SimNet} to calculate the relevance between each interacted item and target item based on the item profile and co-occurrence information:
\begin{equation}
s_{ji} = \textit{SimNet}(\bm{x}_j, \bm{c}_{ji}, \bm{x}_i) = \textit{MLP}\Big(CAT\big(\bm{E}(\bm{x}_j), \bm{E}(\bm{c}_{ji}), \bm{E}(\bm{x}_i)\big) \Big)
\label{eq:sim}
\end{equation}
where $CAT\big(\bm{E}(\bm{x}_j), \bm{E}(\bm{c}_{ji}), \bm{E}(\bm{x}_i)\big) \in \mathcal{R}^{1 \times (2*d_i+d_c)}$ is the concatenation of interacted item embedding, co-occurrence embedding and target item embedding, and $s_{ji}$ represents the relevance between item $j$ and item $i$. $\bm{S}_i = [s_{1i},s_{2i},...,s_{ni}]$ can be considered as a variable dimension representation of target item $i$. We emphasize that
\textit{SimNet} explicitly learns the relevance based on co-occurrence information and side information of items. In this sense, it can be deployed independently for item-to-item retrieval.

After obtaining $\{t_{uj}\}$ and $\{s_{ji}\}$, \baby merges them to get the relevance weight $\textit{PATH}_{uji}$ of each two-hop path as follows:
\begin{equation}
\textit{PATH}_{uji}=\textit{MEG}(t_{uj}, s_{ji})=ln(1+e^{t_{uj}}e^{s_{ji}})
\label{eq:sca}
\end{equation}

\subsection{Direct Net}
We further model the user's general interests in a broader range with another set of user and item embeddings. For example, women are more interested in dresses, while men are more interested in belts. This can be consider as a 1-hop path directly connecting the user to the target item. Hence, we utilize a direct network composed of a user tower and an item tower. Specifically, these two towers go through MLP with Leaky Rectified Linear Units (LeakyReLU) based on user field ($z_u$) and target item field ($x_i$) to output a user representation $\boldsymbol{p}_u \in \mathcal{R}^{1 \times k}$ and an item representation $\boldsymbol{q}_i \in \mathcal{R}^{1 \times k}$, respectively. And then, the relevance weight of the direct path can be formulated as follows:
\begin{equation}
d_{u,i}=\boldsymbol{p}_u\boldsymbol{q}_i^T=\textit{MLP}\big(\bm{E}(z_u) \big)\textit{MLP}\big(\bm{E}(x_i) \big)^T
\label{eq:vec}
\end{equation}
where $d_{u,i}$ is the direct relevance between user $u$ and target item $i$.

\subsection{Bias Net}
Position bias, and many other types of selection biases, are studied and verified to be an important factor in recommender systems~\cite{agarwal2019estimating,zhao2019recommending,chen2020esam}. 
For example, it is common that users are inclined to click items displayed closer to the top of the list, even though it was not the most useful one of the entire corpus. 
To remove selection biases during model training, we train a shallow tower with features contributing to selection bias, such as position feature for position bias, and hour feature for temporal bias. The resultant bias logit $y_{bias}$ is added to the final logit of the main model, as shown in Figure~\ref{fig:pdn_net}. 
Note that, at serving time, the bias net is removed to get an unbiased relevance score.

\subsection{Loss Function}
Whether user $u$ would click target item $i$ can be seen as a binary classification task. Therefore, we merge the relevance weights of $n+1$ paths and bias logit to get the final relevance score between $u$ and $i$, and convert it into user click probability $p_{u,i}$: 
\begin{align}
\hat{y}_{u,i} &=\textit{softplus}(d_{u,i}) + \sum_{j=1}^n\textit{PATH}_{uji}  +  \textit{softplus}(y_{bias})
\label{eq:pdnlogitexp1}\\
 p_{u,i} &= 1 - exp(-\hat{y}_{u,i})\label{eq:pdnlogitexp}
\end{align}
Note that \textit{softplus} function produces the relevance score $\hat{y}_{u,i}$ in the range of $(0,+\infty)$. Hence, we utilize Eq.~\ref{eq:pdnlogitexp} to convert it to a probability value between $0$ and $1$. To train the model, we apply the cross-entropy objective function as:
$l_{u,i} = - \big(y_{u,i} log(p_{u,i}) + (1-y_{u,i})log(1-p_{u,i})\big)$
\label{eq:celoss}
where $y_{u,i}$ is ground-truth label indicating whether the user clicked on the item.

\subsection{Discussion}
\begin{figure}[th]
  	\centering
  	\includegraphics[width=\linewidth]{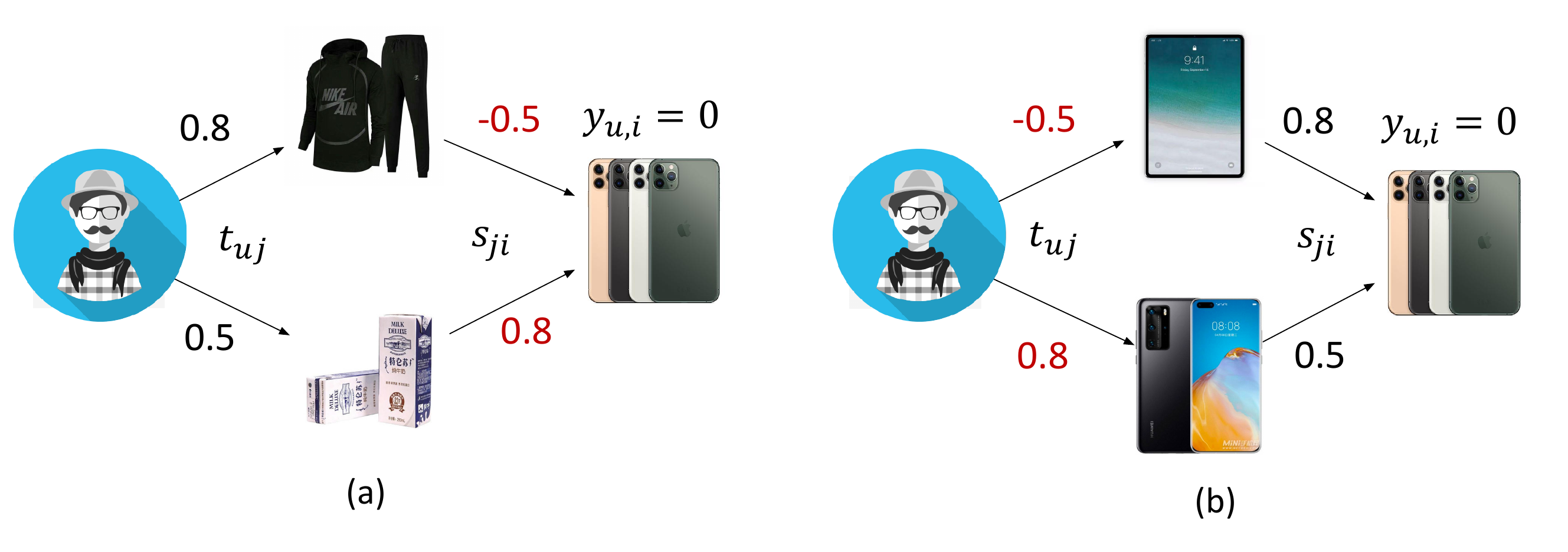}
   	\vspace{-0.5cm}
 	\caption{The bad cases in model training where (a) and (b) are negative sample. To optimize the loss of model, the similarly in (a) and the trigger weight in (b)  have to be negtive.}
 	\label{fig:bad}
\end{figure}

To ensure that the training of \baby can converge to a better optimum, we carefully design the relevance weight for each path. As described above, we utilize $exp(\cdot)$ instead of other activation functions to constrain the output to be positive, i.e., $e^{s_{ji}}$ and $e^{t_{uj}}$, and further, through Eq.~\ref{eq:sca} to achieve the merge of the weights of each 2-hop path. This treatment of constraining the output to be positive is intuitive and fits real-world rationality. Note that the relevance weight would be negative in nature. However, this setting could allow \baby to search the local optimum in a much broader parameter space, which easily leads to overfitting. In Figure~\ref{fig:bad}, We illustrate two bad examples by allowing the relevance weight to be negative. As shown in Figure~\ref{fig:bad}a, when a negative target is connected by two totally irrelevant triggers, \textit{SimNet} may generate a positive relevance weight for one path, and a negative one for the other path. After aggregation, the click probability is still quite low, \ie a perfect match with the ground truth. But it is clear that milk can not have a positive relevance with a mobile phone. Similarly, as shown in Figure~\ref{fig:bad}b, \textit {TrigNet} can also learn a negative preference towards a trigger, mainly to overfit the data.

\begin{figure}[h]
  	\centering
  	\includegraphics[width=0.95\linewidth]{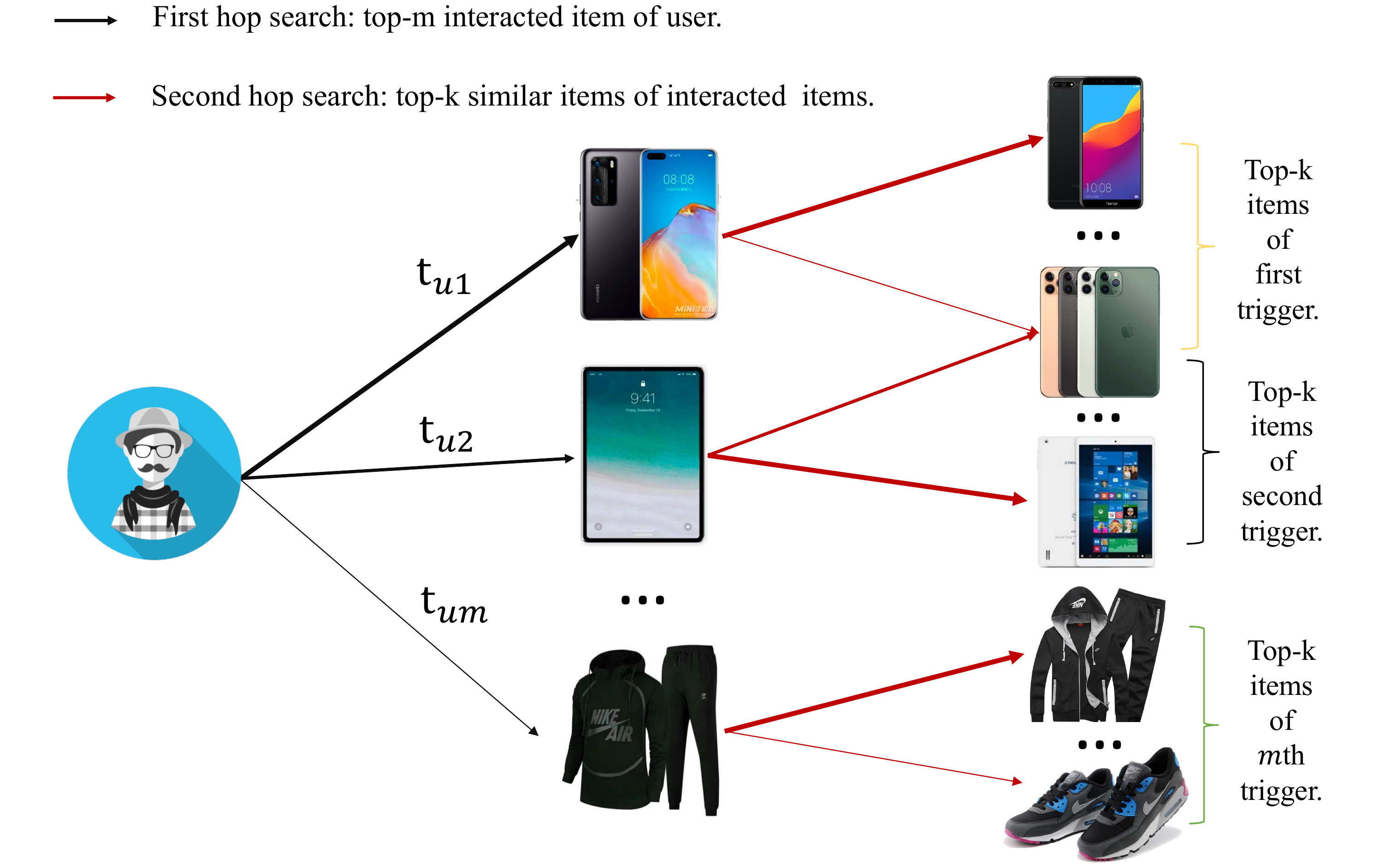}
  	\vspace{-0.2cm}
 	\caption{Online top-$k$ path retrieval with greedy strategy.}
 	\label{fig:pathsearch}
\end{figure}

\begin{figure}[h]
  	\centering
  	\includegraphics[width=0.95\linewidth]{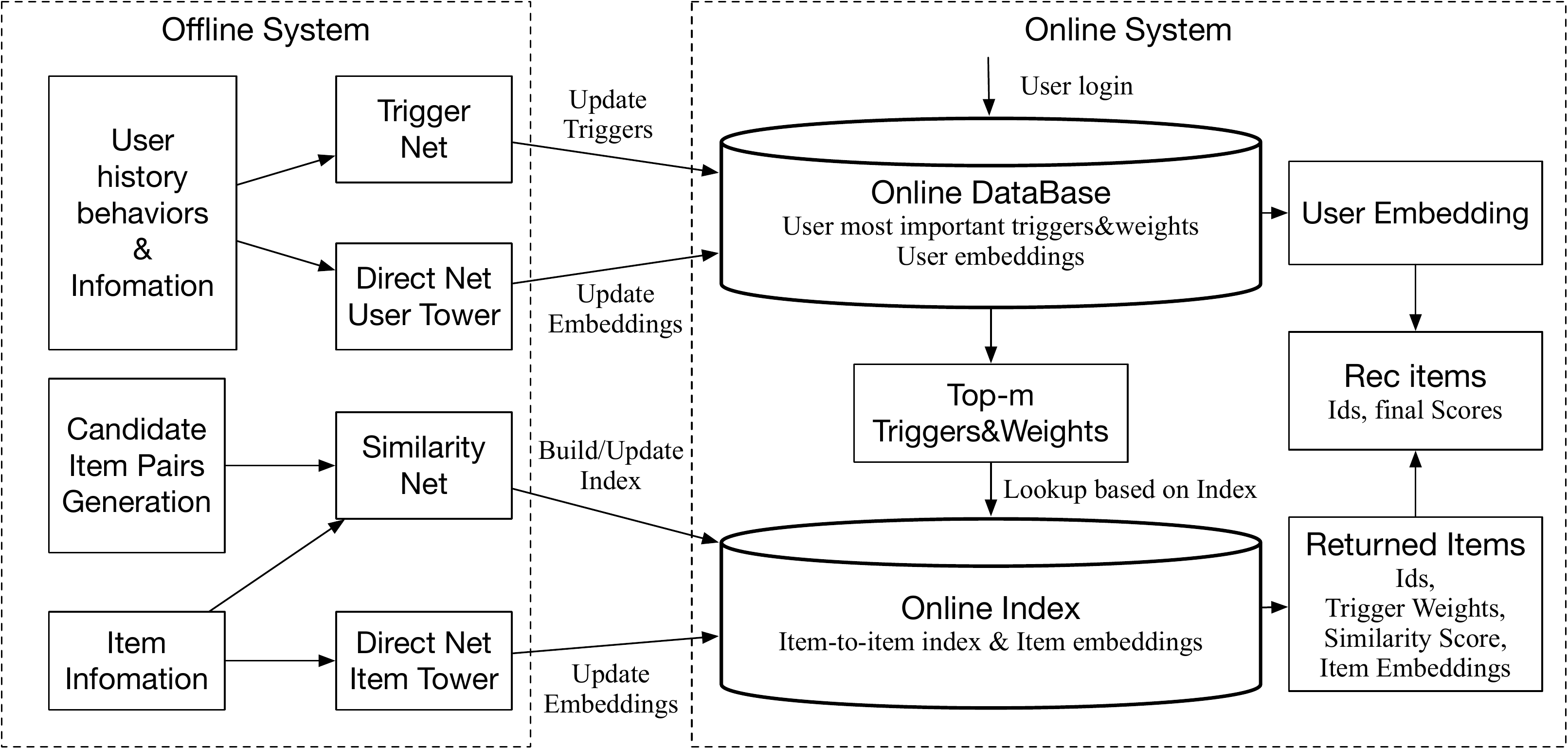}
 	\caption{The framwork of online retrieval with PDN.}
 	\vspace{-0.2cm}
 	\label{fig:pdn}
\end{figure}

\section{System}

In this section, we describe the implementation and deployment of \baby at Taobao for item recommendation in detail. As a user opens Mobile Taobao App, the recommender system firstly retrieves thousands of relevant items for this user from the corpus  containing up to billions of items. Subsequently, every retrieved items are scored by a ranking model and the order list is displayed to the user as the recommendation.

\subsection{Onling Retrieval}
As aforementioned, billions or even trillions of items need to be processed in the matching stage. To satisfy the online real-time services, it is not possible to utilize \baby to score all available items. Based on architecture of \baby, it is intuitive that  the larger path relevance weight, the more likely the user would be interested in the item. Therefore, the matching problem can be regarded as retrieving the item node with the larger path weights in the user's 2-hop neighborhood.
As shown in Figure~\ref{fig:pathsearch}, we implement online real-time top-$K$ nearest neighbor retrieval in the form of path retrieval via a greedy strategy. 
Specifically, we decompose the path retrieval into two parts: 
(a) top-$m$ important trigger search (the first hop) using the \textit{TrigNet},
(b) top-$k$ item search for each top-$m$ trigger (the second hop) based on the index generated by \textit{SimNet}.
For \textit{TrigNet}, we deploy the model in a real-time online service for user trigger scoring. For \textit{SimNet}, we use the model to compute and index item relevance in offline fashion. In detail, the online retrieval of PDN can be summarized as:
\begin{itemize}
\item  \textbf{Index generation (Step 1):} Based on \textit{SimNet}, the $k$ most similar items of each item in the corpus are generated offline and stored with the relevance score $s_{ji}$ in the database. The details can be seen in Section~\ref{sec:index}

\item \textbf{Trigger extraction (Step 2):} When the user opens Mobile Taobao App, we take out all the items that the user has interacted with and utilize \textit{TrigNet} to score all her triggers $t_{uj}$ and return top-$m$ triggers.

\item \textbf{Top-$K$ retrieval (Step 3):} We start with these top-$m$ triggers to query the database and obtain $m \times k$ candidate items in total. Without considering the bias feature, the top-$K$ items are returned for recommendation as follows. 
\begin{equation}
\hat{s}_{u,i} =\textit{softplus}(d_{u,i}) + \sum_{j=1}^m\textit{softolus}(t_{uj}+s_{ji})
\label{eq:sco}
\end{equation}
\end{itemize}
Note that $\boldsymbol{p}_u$ and $\boldsymbol{q}_i$ are static representations. Hence, these two representations can also be inferred offline and stored in the database. When performing online service, they can be queried directly based on user or item id.

\subsection{Index Generation} 
\label{sec:index}
For industrial systems with large item corpus, we need to compress the dense matrix of item relevance, \ie $\mathcal{R}^{N \times N}\rightarrow\mathcal{R}^{N \times k}$, to reduce index construction time and storage cost. Even this process can be finished offline, it is too cost to calculate the relevance of all $N \times N$ item pairs, so we first reduce $N \times N$ to $N \times \hat{k}$ based on candidate generation, where $\hat{k}$ is an order of magnitude greater than $k$.
The compression operation mainly consists of three steps:
\begin{itemize}
\item \textbf{Candidate item pair generation.} We mainly generate item pairs from two strategies, one is based on co-occurrence information, such as items that are clicked in the same session, the other is based on the profile information of items, such as items of the same brand. In this sense, the item pairs which have been not co-occurred before but have some similar property can also be considered as candidates.
\item \textbf{Candidate item pair ranking.} We extract the relevance score $s_{ji}$ of each item pair by using \textit{SimNet}.
\item \textbf{Index building.} For each item, we take the top-$k$ similar items to form $N \times k$ item pairs, and store them with $s_{ji}$ in the database. 
\end{itemize}
It is worth mentioning that \textit{SimNet} can be used independently for indexing, so it can be used also for item-to-item retrieval.

\section{Experiements} 

In this section, we evaluate \baby against the existing state-of-the-art solutions with both offline and online settings, including ablation study, model analysis and case study.

\begin{table}[ht]
  \vspace{-0.2cm}
  \caption{Statistics of experimental datasets.}
  \vspace{-0.3cm}
  \label{tab:data}
  \begin{tabular}{cccc}
    \toprule
    \textbf{Dataset}&\textbf{User\#}&\textbf{Item\#}&\textbf{Interaction\#}\\
    \midrule
    MovieLens & 6,040&3,706&1,000,209\\
    Pinterest & 55,187&9,916&1,500,809\\
    Amazon Books &351,356&393,801&6,271,511\\
  \bottomrule
  \end{tabular}
  \vspace{-0.5cm}
\end{table}

\begin{table*}[ht]
  \centering
  \caption{HR and NDCG of different methods on the three datasets. The best results are highlighted in boldface. The improvements over the comparing methods are statistically significance at $0.05$ level. 'Personalise' means that the output of the score considers the user information, e.g., the output of PDN and EBR-based method. It is worth noting that CF can't get personalized scores, which is represented by '-'. 'Item to Item' refers to retrieval based on the relevance between interacted and target items. Specifically, the acquisition strategy of item similarity is the output of item-based CF, the output of \textit{SimNet} or the inner product between item features extracted from EBR methods.}
    \vspace{-0.2cm}
  \label{tab:res}\resizebox{\linewidth}{!}{
    \begin{tabular}{cccccccccccccc}
   \toprule
    \multirow{3}{*}{Group}&\multirow{3}{*}{Method}&
    \multicolumn{4}{c}{MovieLens}&\multicolumn{4}{c}{Pinterest}&\multicolumn{4}{c}{Amazon Books}\cr
    \cmidrule(lr){3-6} \cmidrule(lr){7-10}\cmidrule(lr){11-14}
    &&\multicolumn{2}{c}{Personalise}&\multicolumn{2}{c}{Item to Item}&\multicolumn{2}{c}{Personalise}&\multicolumn{2}{c}{Item to Item}&\multicolumn{2}{c}{Personalise}&\multicolumn{2}{c}{Item to Item}\cr
    \cmidrule(lr){3-4} \cmidrule(lr){5-6}\cmidrule(lr){7-8} \cmidrule(lr){9-10} \cmidrule(lr){11-12} \cmidrule(lr){13-14}
    &&HR@10&NDCG@10&HR@10&NDCG@10&HR@10&NDCG@10&HR@10&NDCG@10&HR@10&NDCG@10&HR@10&NDCG@10\cr
    \midrule
    \multirow{3}{*}{Two-tower}&DSSM \cite{DSSM}&0.4699&0.2603&0.2243&0.1034&0.2730&0.1394&0.1647&0.0759&0.6433&0.4163&0.1981&0.0894\cr
    &Youtube DNN \cite{YoutBednn}&0.6187&0.3579&0.4362&0.2324&0.7585&0.4351&0.6409&0.3028&0.6539&0.4370&0.2398&0.0955\cr
    &BST\cite{chen2019behavior}&0.6316&0.3603&0.0561&0.0277&0.8176&0.4955&0.5371&0.2428&0.6923&0.4403&0.1019&0.0574\cr
    \midrule
     \multirow{2}{*}{Item-based CF}&PCF \cite{itemcf} &-&-&0.4033&0.2132&-&-&0.2800&0.1470&-&-&0.0547&0.0203\cr
    &SLIM \cite{ning2011slim}&-&-&0.4400&0.2337&-&-&0.6067&0.3323&-&-&0.1621&0.0679 \cr
    \midrule
    ranking model&DIN \cite{din}&0.6454&0.3767&0.1662&0.0918&0.8185&0.5051&0.4370&0.2059&0.6973&0.4698&0.1325&0.0612 \cr
    \midrule
    \multirow{3}{*}{Incorporated}&PDN w/o bias net &0.6323&0.3401&0.5080&0.2795&0.8015&0.4814&0.7654&0.4345&0.6869&0.4520&0.2866&0.1546\cr
    &PDN w/o direct net &0.6642&0.3930&0.5124&0.2813&0.8123&0.5286&0.7703&0.4397&0.7012&0.4729&0.2962&0.1613\cr
    &PDN  &\textbf{0.6770}&\textbf{0.4071}&\textbf{0.5152}&\textbf{0.2859}&\textbf{0.8283}&\textbf{0.5358}&\textbf{0.7911}&\textbf{0.4613}&\textbf{0.7019}&\textbf{0.4735}&\textbf{0.3505}&\textbf{0.2049}\cr
    \bottomrule
    \end{tabular}}
    \vspace{-0.3cm}
\end{table*}

\subsection{Offline Evaluation} \label{Sec:E1}
\subsubsection{Datasets and Evaluation Protocol.}
We experiment with three publicly real-world datasets: MovieLens \footnote{http://grouplens.org/datasets/movielens/1m/}, 
Pinterest~\cite{geng2015learning}, and Amazon books~\cite{he2016ups} for performance evaluation. The statistics of the three datasets are summarized in Table~\ref{tab:data}. 
Following the settings of~\cite{he2017neural}, we filter these datasets in the same way that retained only users with at least $20$ interactions. More details on this preprocessing for the three datasets have been elaborated in~\cite{he2017neural}, so we do not restate here. 

To evaluate the performance of \baby, we adopt the leave-one-out evaluation protocol~\cite{bayer2017generic,he2016fast}. For each user, the last interaction is used as the target item, while the previous interaction items are collected as the user behaviors. 
Specifically, we followed the common strategy in~\cite{elkahky2015multi,koren2008factorization}. That is, all negative items are utilized in each test case, and the target items are ranked among these items. Hit Ratio (HR)~\cite{deshpande2004item} and Normalized Discounted Cumulative Gain (NDCG)~\cite{he2015trirank} are adopted as the performance metrics. Here HR can be interpreted as a recall-based measure and NDCG is a ranking-based measure.

\subsubsection{Comparing Methods.}
We compare \baby with the following two-tower methods and conventional item-based CF methods.
The two-tower methods are introduced as follows:
\begin{itemize}
\item \textbf{DSSM \cite{DSSM}.} 
DSSM employs embeddings to represent users and items.  
The relevance score is calculated based on the inner product between the user representation and the item representation.

\item \textbf{Youtube DNN \cite{YoutBednn}.} 
Youtube DNN utilizes the user's interaction sequence to derive user representations for item recommendation. 
It treats each item in the user's historical behaviors equally and adopts average pooling to extract a user's interest.
To ensure fair comparison, hyperparameter tuning is conducted by a grid search, and each method is tested with the best hyperparameters.

\item \textbf{BST \cite{chen2019behavior}.}
BST extends the Youtube DNN, 
by utilizing the transformer layers to capture the user's short-term interest over the behavior sequence.
We use the inner poduct of user and item representations instead of MLP.
\end{itemize}
The item-based CF  methods are introduced as follows:
\begin{itemize}
\item \textbf{Pearson-based CF (PCF) \cite{itemcf}.} This is the standard item-based CF, which estimates item relevance based on Pearson coefficient. 

\item \textbf{SLIM \cite{ning2011slim}.} SLIM learns item relevance by minimizing the loss between the original user-item interaction matrix and the reconstructed one from the item-based CF model. 
\end{itemize}
The ranking  method is introduced as follows:
\begin{itemize}
\item \textbf{DIN \cite{din}.} Deep interest network takes target attention to extract the relationship between user's interacted sequence and target item.
\end{itemize}

\subsubsection{Results and Discussion} 
Table~\ref{tab:res} shows the experimental results on three public datasets in terms of HR@10 and NDCG@10. All experiments are repeated $5$ times and the average results are reported. Clearly, PDN outperforms all comparison methods under most datasets. We can make the following observations. (1) For personalized retrieval, DSSM performs worst among two-tower methods, which indicates that capturing user interest based on the user behavior is critical for recommender systems. The performance of BST is better than that of YouTube DNN, due to the fact that the transformation layer extracts users' interests by considering the sequential information in their behaviors. PDN achieves the best performance, mainly because these two-tower methods represent each user by one fixed-dimension vector, \ie a bottleneck for modeling diverse interests. In contrast, PDN utilizes \textit{TrigNet} to extract multi-interest user representation and each dimension describes the user's interest in an interacted item in a fine-grained way. \textit{SimNet} is also effective in deriving the similarity between an interacted item and the target item (ref. Figure~\ref{fig:path}). Hence, a more accurate relevance can be estimated by considering the potential interest of the user towards the target item. 

(2) When deploying online, we adopt SimNet instead of item-based CF to estimate item similarity for item-to-item retrieval. Therefore, we conducted a set of comparison with an item-to-item strategy (\ie the columns with ``Item to Item'' in Table~\ref{tab:res}). SimNet performs best among all methods. 
The reason is that SimNet explicitly optimizes the similarity between items by integrating the item profile used by the two-tower methods and co-occurrence information used by item-based CF  based on deep neural networks, 
which utilizes more information to solve sparsity problem encountered by CF.
(3) PDN performs better than DIN. We believe that DIN ignores users' attention to interacted items, while PDN considers both user attention and item attention for better personalized recommendation. (4) Based on ablation study, PDN yields the best performance, confirming that each component contributes to the final results.

\subsection{Online Experiments}
\label{Sec:E2}

\subsubsection{A/B Tests.} 
\begin{figure}[h]
    \centering
    \includegraphics[width=0.8\linewidth]{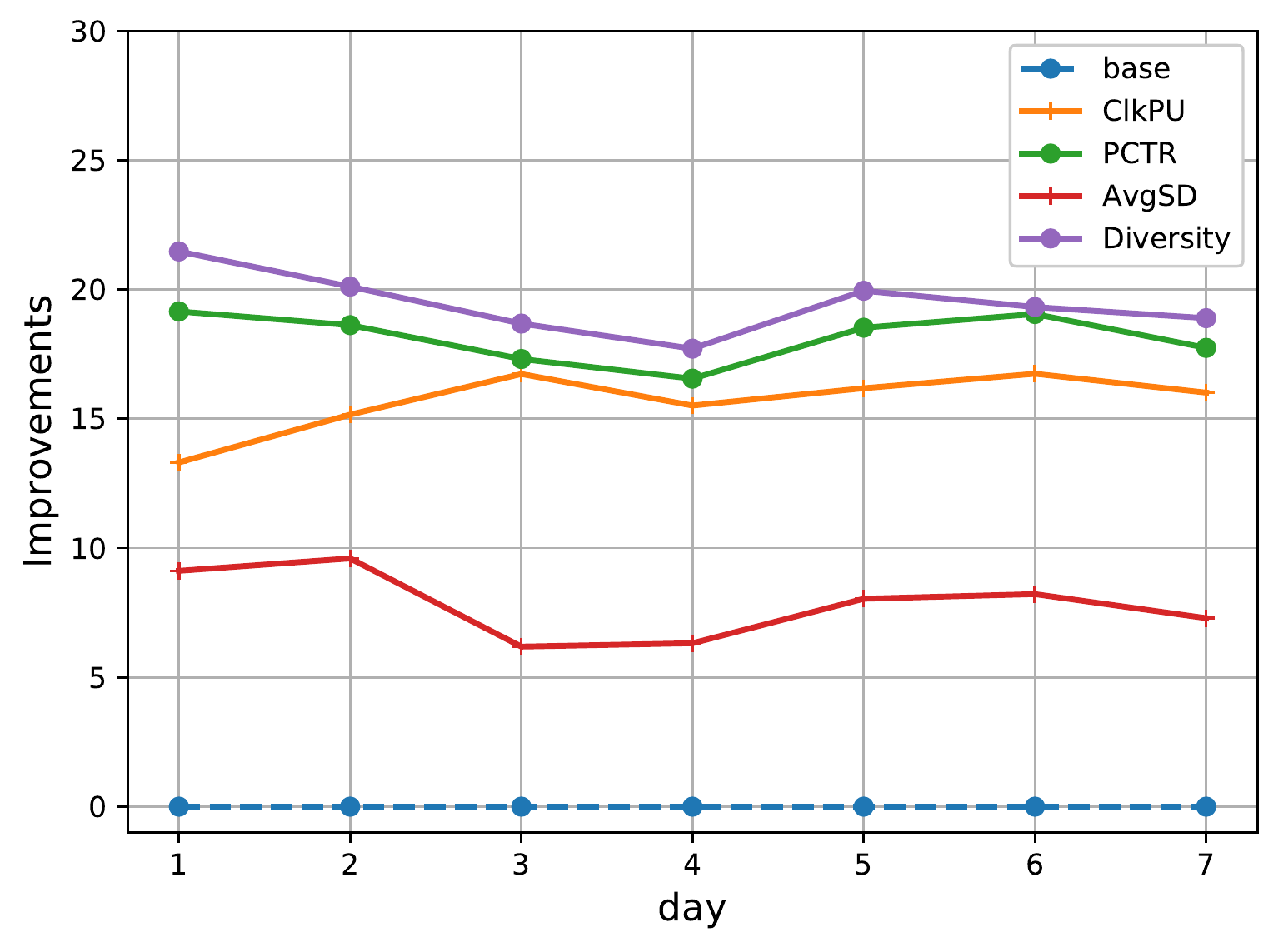}
    \vspace{-0.3cm}
    \caption{PDN online improvements in one week.}
    \label{fig:online}
    \vspace{-0.4cm}
\end{figure}

Beyond offline studies, we conduct online A/B experiments by deploying our method in the recommender system of Taobao for two weeks. 
In the control setup (\ie Baseline), it includes all matching strategies in our current production system. In the variation experiment setup,  
we applies \textit{SimNet} instead of the item-based method for inverted-index building. 
For fair comparison, the same ranking component and business logic are applied on the top of both matching stages. Table~\ref{tab:ab} and Figure~\ref{fig:online} summarize the experimental results. We can find that the proposed \baby improves the e-commerce recommender system for all core business metrics, including page click-through rate (PCTR), user click-through rate (UCTR), clicks per user (ClkPU), average session duration (avgSD), as well as diversity\footnote{Diversity represents the proportion of category coverage.}, which are considered to be good indicators of recommendation satisfaction. Especially, personalized diversity is normally hard for existing production systems, while PDN increases the diversity of recommended items by a large margin, which indicates PDN can capture the diverse interests of users with \textit{TrigNet} and \textit{SimNet} to improve the overall user experience.
Besides, we deploy \textit{SimNet}  to verify it can be used independently to build index instead of item-based CF for item-to-item retrieval, which also gets a  perform gains. Note that, the metrics online are reported with relative improvement, \ie $(metric_{\baby}-metric_{base})/metric_{base}$.

\begin{table}[h]
  \centering
  \caption{Online A/B test improvements in Taobao (Observed over two week).}
  \vspace{-0.2cm}
  \label{tab:ab}
    \begin{tabular}{cccccc}
    \toprule
    Method&PCTR&ClkPU&AvgSD&Diversity\cr
    \midrule
    Baseline&$+$0.0\%&$+$0.0\%&$+$0.0\%&$+$0.0\%\cr
    \textit{SimNet}&$+$9.25\%&$+$5.43\%&$+$4.08\%&$+$14.68\%\cr
    PDN&\textbf{$+$18.04\%}&\textbf{$+$15.38\%}&\textbf{$+$7.87\%}&\textbf{$+$19.60\%}\cr
    \bottomrule
    \end{tabular}
    \vspace{-0.7cm}
\end{table}

\subsubsection{Online Efficiency.} Here, we report the efficiency of online serving. Specifically, the overall latency from requests to candidate generation can be done within 6.75 milliseconds, \ie the queries per second (QPS) is in the thousand level, This is comparable to the retrieval with the standard inverted index. 
Based on the huge performance gain and low latency, we have deployed PDN online to serve the matching stage of recommendation in Taobao.


\subsection{Case Study} \label{Sec:E3}

\begin{table}[h]
  \centering
  \caption{Impact of user behavior sequence length ($n$) on HR@300 at Taobao offline logs. Percentages in the brackets indicate the relative improvements over BST.}
  \vspace{-0.2cm}
  \label{tab:sqlen}\resizebox{\linewidth}{!}{
    \begin{tabular}{cccccc}
   \toprule
    Method&$n$<=15&15<$n$<=30&30<$n$<=45&$n$>45&all\cr
    \midrule
    PCF~\cite{itemcf}&0.066 (-46.3\%)&0.110 (-59.1\%)&0.131 (-31.4\%)&0.141 (-41.8\%)&0.120 \cr
    BST~\cite{chen2019behavior}&0.123&0.175&0.191&0.200&0.180\cr
    \bottomrule
    PDN&\textbf{0.263 (+113.8\%)}&\textbf{0.326 (+86.3\%) }&\textbf{0.329 (+72.3\%)}&\textbf{0.295 (+47.5\%)}&0.297\cr
    \bottomrule
    \end{tabular}}
    \vspace{-0.5cm}
\end{table}

\begin{figure}[h]
    \centering
    \includegraphics[width=0.8\linewidth]{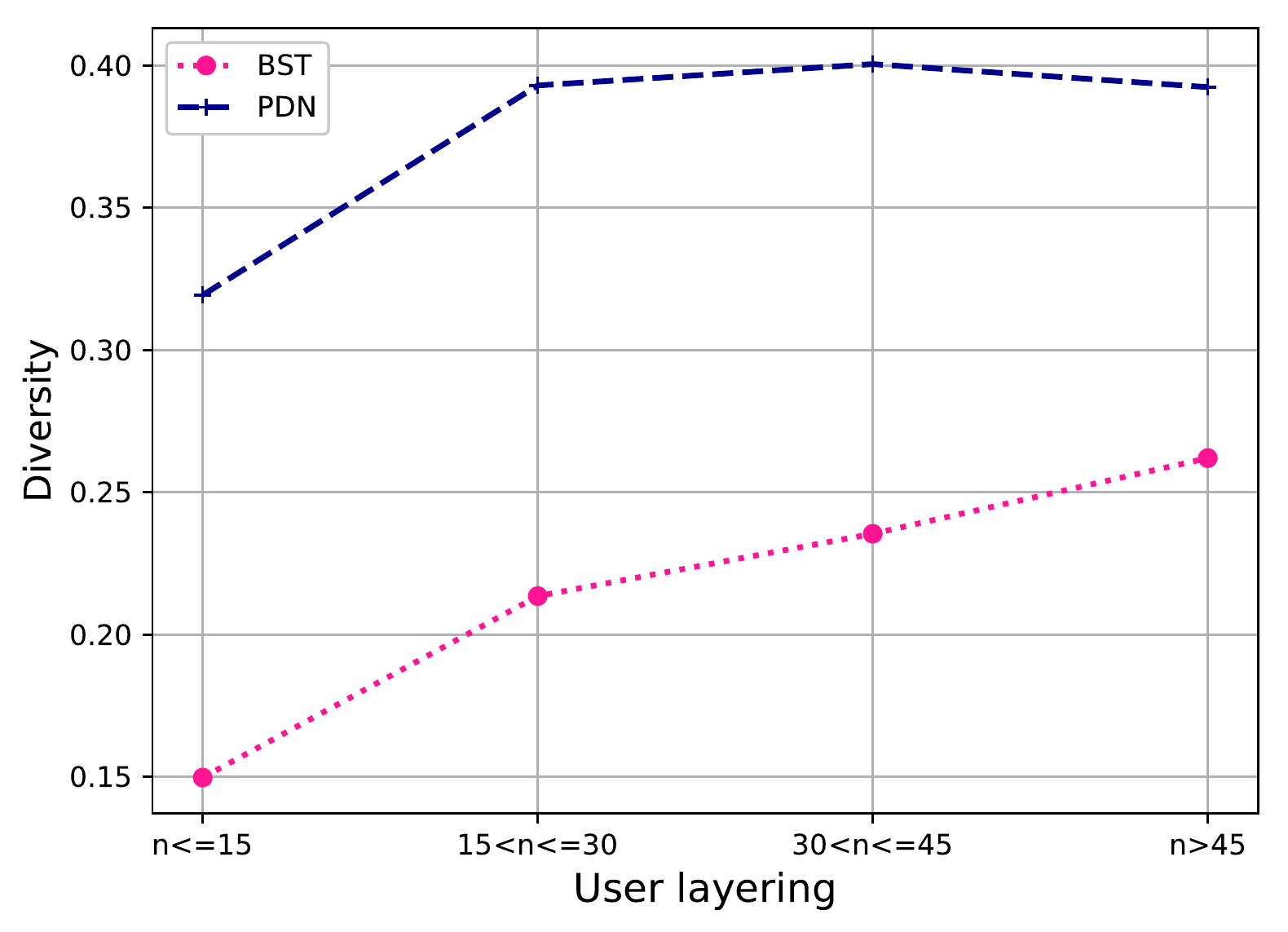}
    \vspace{-0.3cm}
    \caption{Diversity w.r.t. user behavior sequence length.}
    \label{fig:diversity}
    \vspace{-0.3cm}
\end{figure}

\begin{figure*}[h]\label{case}
    \centering
  \begin{subfigure}[b]{0.25\linewidth}
    \includegraphics[width=\linewidth]{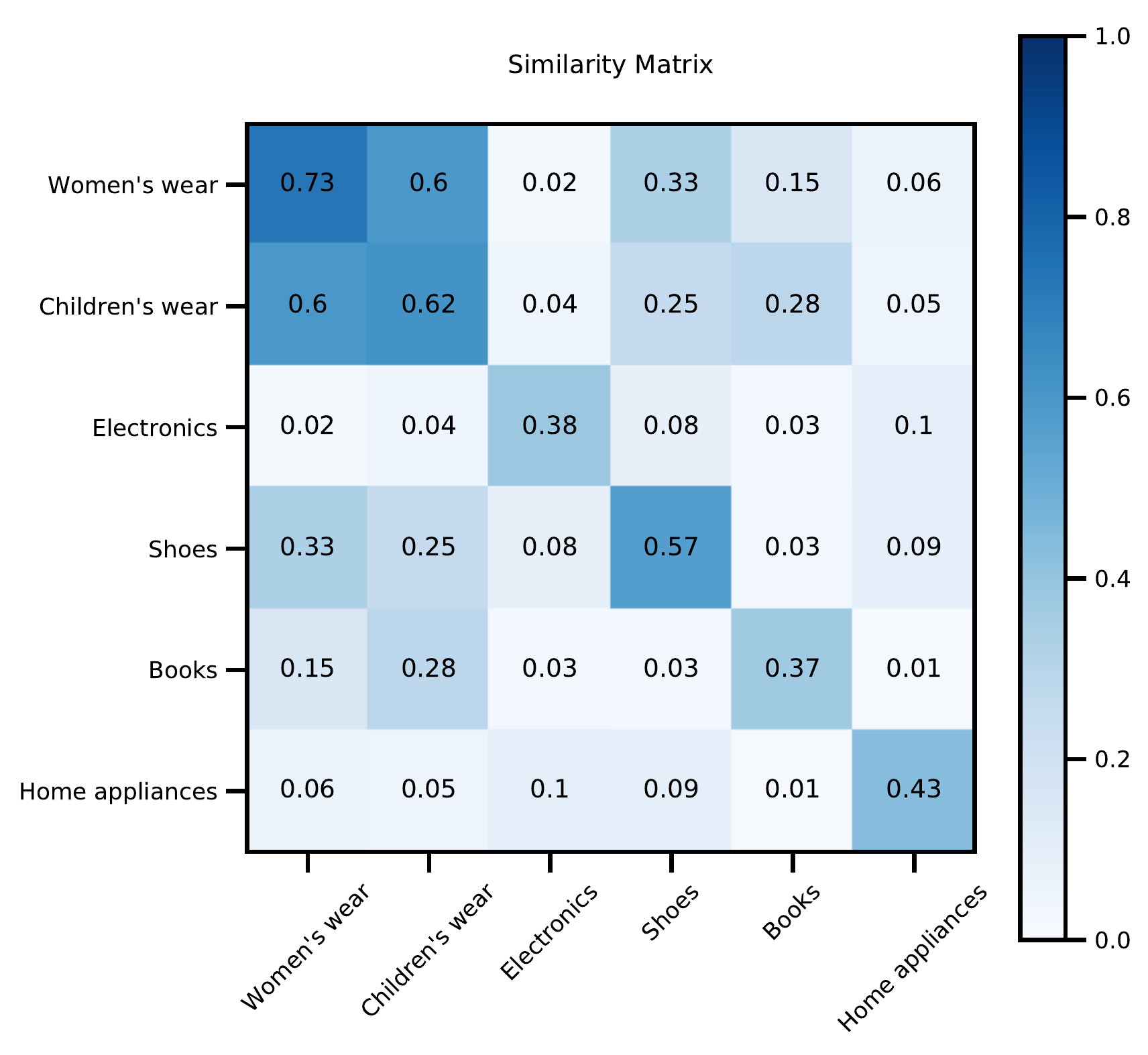}
    \caption{ }
    \label{fig:sim_case2}
  \end{subfigure}
   \begin{subfigure}[b]{0.325\linewidth}
    \includegraphics[width=\linewidth]{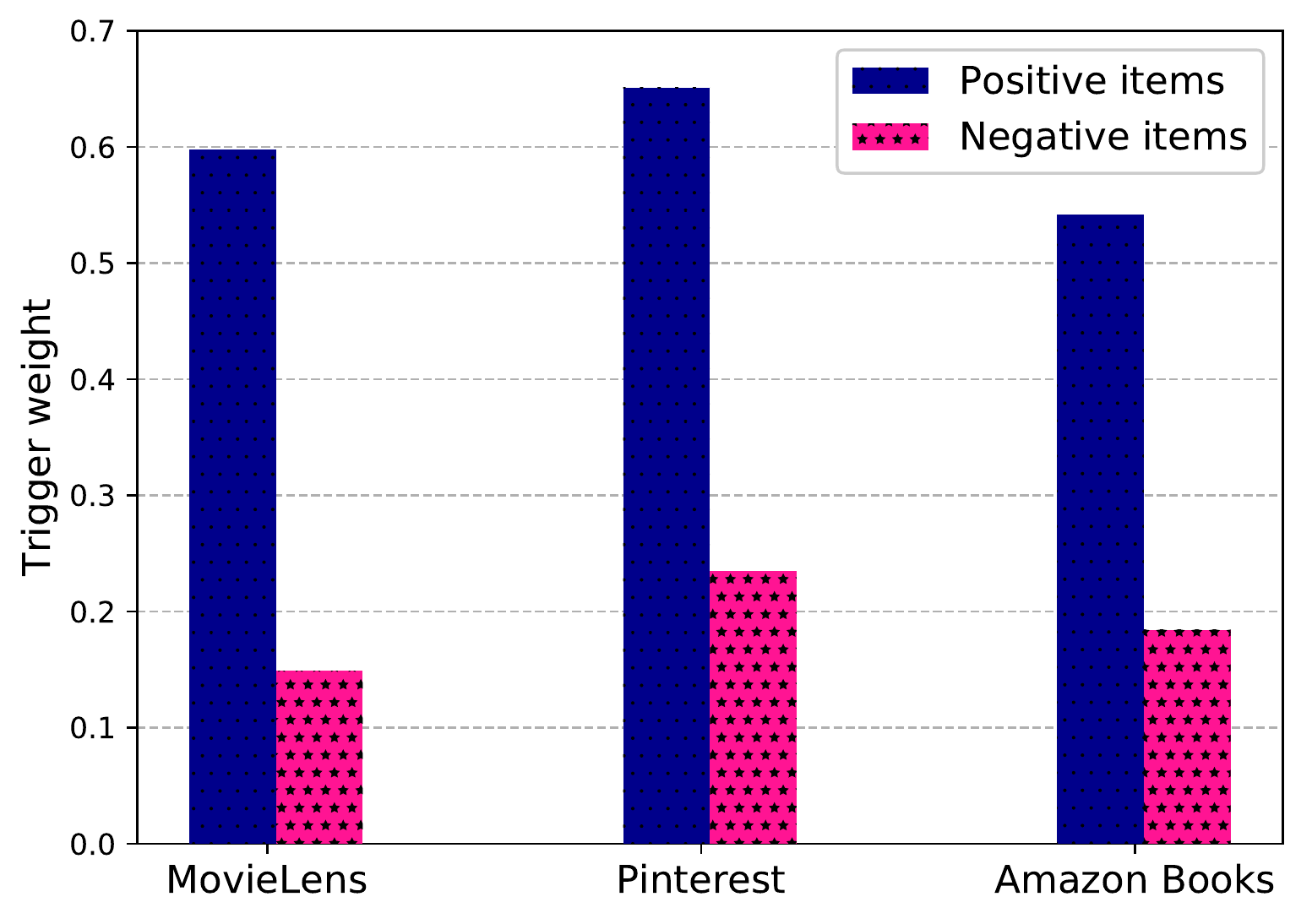}
    \caption{ }
 \label{fig:tri_case}
  \end{subfigure}
    \begin{subfigure}[b]{0.3505\linewidth}
    \includegraphics[width=\linewidth]{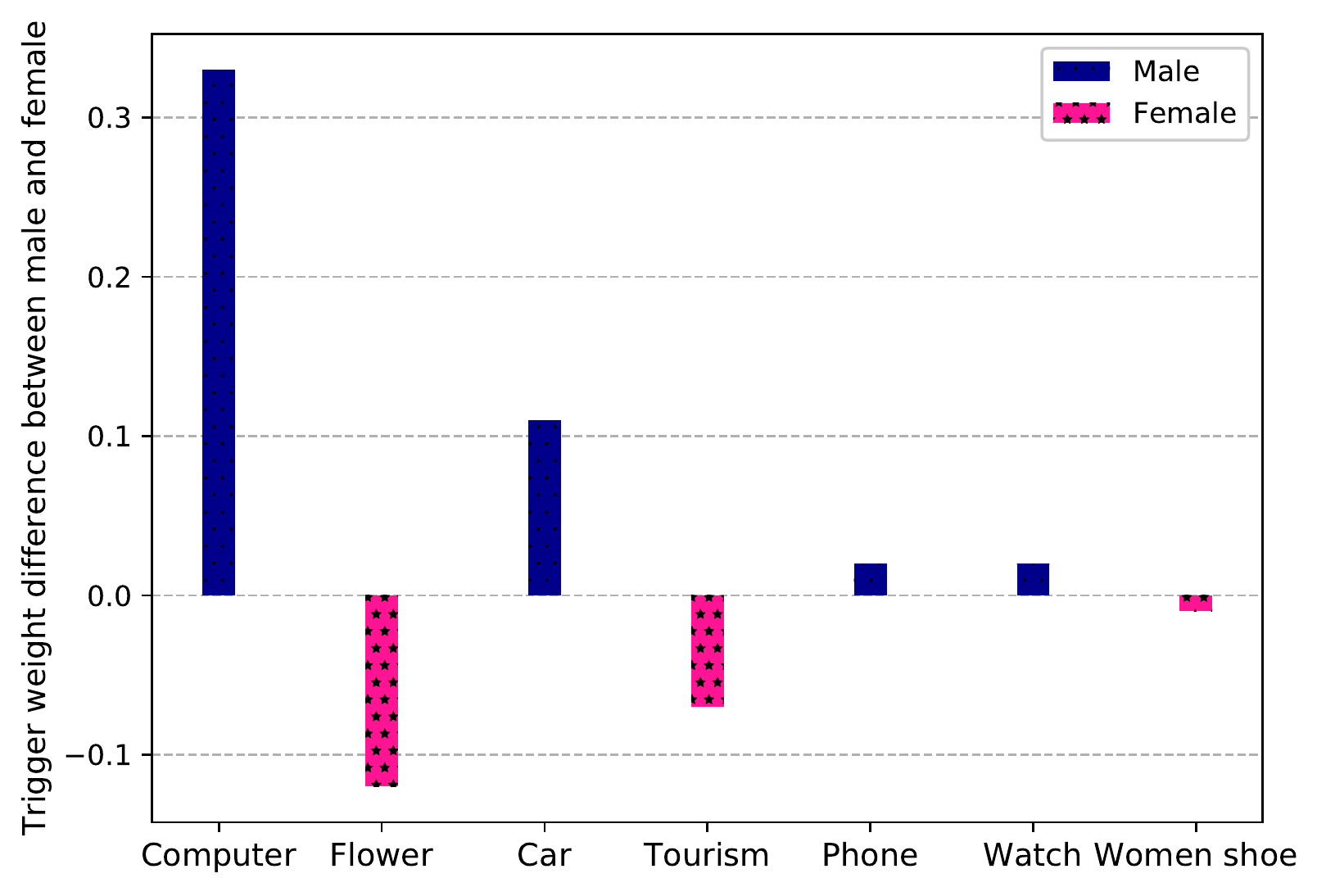}
    \caption{ }
 \label{fig:tri_per}
  \end{subfigure}
\caption{(a) Item-to-item similarity matrix based on \textit{SimNet} at online serving; (b) User-to-item interests based on \textit{TrigNet} on public datasets; (c) User group-to-category interests based on \textit{TrigNet} at online serving.}
\vspace{-0.2cm}
\end{figure*}

\begin{figure}
    \centering
    \includegraphics[width=\linewidth]{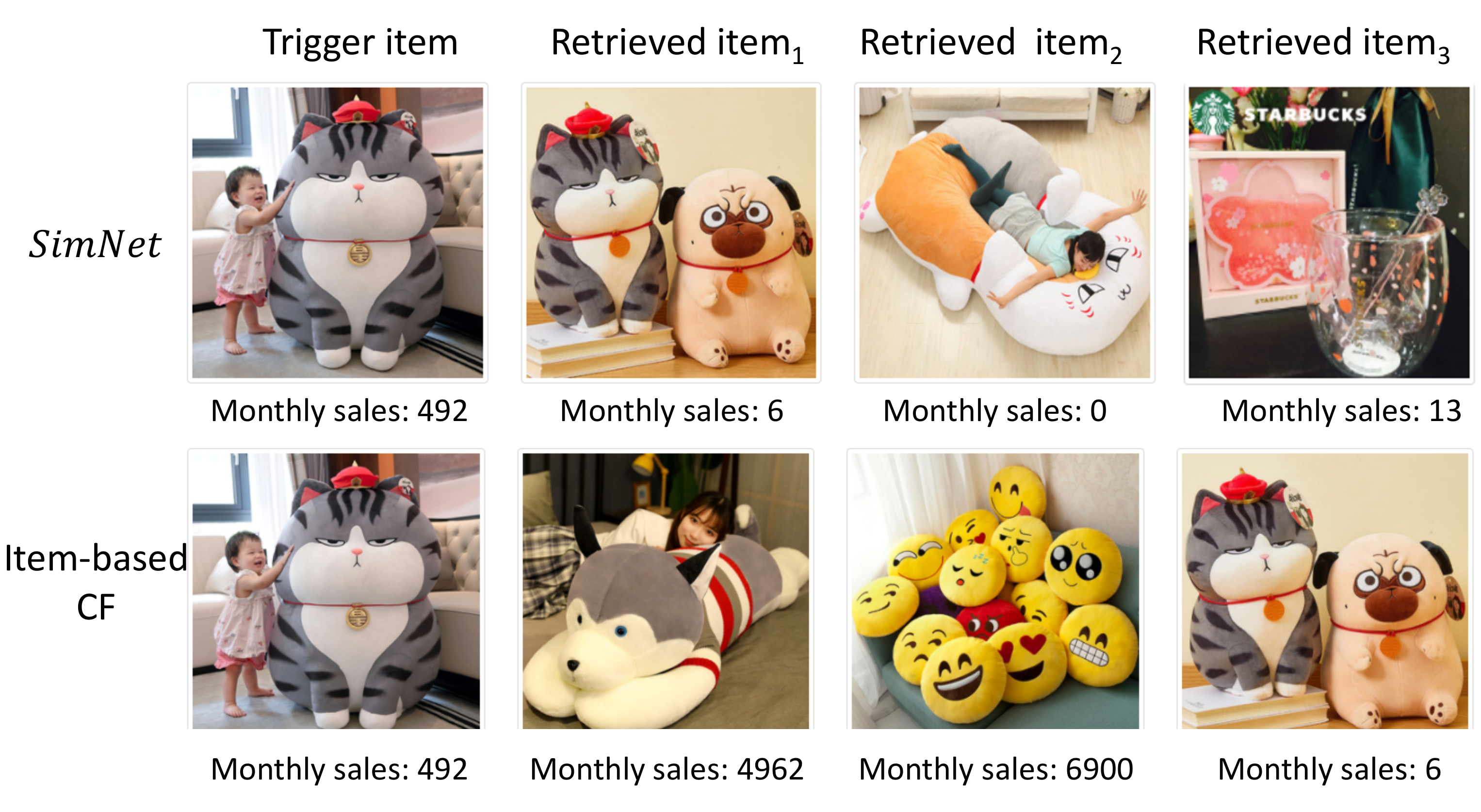}
  \caption{Comparison of item-to-item retrieval results between \textit{SimNet} and item-based CF based on trigger item. The similarity decreases from left to right.
}
  \label{fig:sim_case}
  \vspace{-0.3cm}
\end{figure}

\subsubsection{Analysis of User Behavior Sequence Length.} Based on the model trained with Taobao offline log, we investigate the impact of user behavior sequence length $n$ on performance. As shown in Table~\ref{tab:sqlen}, we group users based on $n$, and found that the smaller the $n$ is, the larger the gain is. This result shows that PDN has stronger robustness to  $n$, and it can obtain superior performance even in the case of sparse user behavior sequence. As shown in Figure~\ref{fig:diversity}, with the increase of $n$, the diversity of candidates generated by PDN gradually increases, and is better than that of BST by a large margin. These results show that PDN can capture the diverse interests of users in a fine-grained way to improve the user experience.

\subsubsection{Diversity and Accuracy of \textit{SimNet}.}
To verify the effectiveness of \textit{SimNet}, we utilize the same trigger item to perform item-to-item retrieval using the similarity provided by \textit{SimNet} and item-based CF, respectively. As shown in Figure~\ref{fig:sim_case}, retrieval through cat-shaped dolls, \textit{SimNet} returns similar cat-shaped dolls and cups printed with cat paw patterns, while item-based CF returns dolls with high monthly sales but not related to cats, \eg the dog-shaped dolls. This result indicates that \textit{SimNet} can retrieve more relevant and diverse results without being disturbed by the number of interactions, while item-based CF would suffer from ``Matthew Effect''~\cite{krishnan2018adversarial}, which introduces perference bias towards popular items. In other words, item-based CF only calculates similarity based on item co-occurrence information, while our method additionally introduces more information, \eg item profile, user profile, user behavior, for more accurate item similarity estimation. To further verify the reliability of item similarity based on \textit{SimNet}, we randomly select $1,000$ item pairs based on each category pair and calculate their average similarity from the inverted index built by \textit{SimNet}. As shown in Figure~\ref{fig:sim_case2}, we find that items belonging to the same category have a higher similarity. Besides, the similarity of related categories is higher than that of unrelated categories. For example, both the pair of women's wear and children's wear, and the pair of electronics and home appliances, have higher similarity than other combinations.

\subsubsection{Personalization and Effectiveness of TrigNet.} On the public datasets, we exploit the user's interest in items based on \textit{TrigNet}. As shown in Figure~\ref{fig:tri_case},  the user-to-item trigger weights are averaged on positive and negative items respectively. It is obvious that the weights on positive items are higher than that on negative items. This result indicates the reasonable of \textit{TrigNet}. To verify the personalization ability of \textit{TrigNet}, we randomly selected $1,000$ males and $1,000$ females, and randomly selected $500$ items from several categories, and then apply \textit{TrigNet} to score all user-item pairs and get the average trigger weight difference between male and female. As shown in Figure~\ref{fig:tri_per}, we can find that \textit{TrigNet} treats gender differently to reflect the user characteristics precisely. For example, male prefer computer, car, and watch, while female prefer flower, tourism and women shoes.

\section{Conclusion}
In this paper, a novel model named Path-based Deep Network is proposed for candidate item generation. The proposed \baby establishes 2-hop paths from the user to the target item, leading to better personalized yet diverse item retrieval. To the best of our knowledge, this is the first work to build a retrieval architecture based on a 2-hop graph. Besides, we present how to apply this architecture to achieve online retrieval in low latency based on the proposed path retrieval and a greedy strategy. The offline experiments over three real-world datasets are conducted to demonstrate that \baby outperforms existing SOTA alternatives. Moreover, our online experiments further indicates that \baby can improve retrieval quality in terms of five industry metrics. Lastly, \baby has been successfully deployed in our online recommender system in Taobao.

\begin{acks}
    Chenliang Li's work was supported by National Natural Science Foundation of China (No.~61872278).
\end{acks}
\bibliographystyle{ACM-Reference-Format}
\bibliography{sample-base}

\end{document}